\documentclass[12pt,preprint]{aastex}

\newcommand{\Lsun}{{L$_{\odot}$}}
\newcommand{\hii}{{H\footnotesize{II} }}
\def\fmag{\hbox{$.\!\!^{\rm m}$}}

\shorttitle{Star-forming Regions in the SMC}
\shortauthors{Livanou et al.}

\begin{document}


\title{Star-forming Regions in the Small Magellanic Cloud \\
    Multi-wavelength Properties of Stellar Complexes}


\author{E. Livanou and I. Gonidakis}
\affil{Department of Astrophysics Astronomy \& Mechanics,
 Faculty of Physics, University of Athens, GR-157 83 Athens, Greece}

\author{E. Kontizas}
\affil{Institute for Astronomy and Astrophysics, National
 Observatory of Athens, P.O. Box 20048, GR-118 10 Athens, Greece}

\author{U. Klein}
\affil{Argelander-Institut der Universit\"at Bonn, Auf dem H\"ugel 
71, 53121, Bonn, Germany}

\author{M. Kontizas}
\affil{Department of Astrophysics Astronomy \& Mechanics,
 Faculty of Physics, University of Athens, GR-157 83 Athens, Greece}

\author{D. Kester}
\affil{SRON, Netherlands Institute for Space Research, P. O. Box 800, 9700 AV Groningen}

\author{Y. Fukui and N. Mizuno}
\affil{Department of Astrophysics, Nagoya University, Chikusa-ku, Nagoya
 464-8602}

\and

\author{P. Tsalmantza}
\affil{Department of Astrophysics Astronomy \& Mechanics,
 Faculty of Physics, University of Athens, GR-157 83 Athens, Greece}




\begin{abstract}
We trace the star formation regions in the SMC and study their properties.
The size and spatial distribution of these regions is found to support the 
hierarchical scenario of star formation, whereas, the evaluation of their 
intensity, contributes to the understanding of the various stages of star formation.
Their connection to the LMC-SMC close encounter, about $(0.9-2)~\times~10^{8}$ 
years ago, is investigated as well.
The SMC, being almost edge-on, does not easily reveal these areas, as is the 
case with the LMC. However, a study through multi-wavelength images such as 
optical, IR and radio has been proved very useful. A selection of areas, 
with enhanced 60 and 100-$\mu$m infrared flux and emission in all IRAS bands, 
identifies the star forming regions. 
All of the identified regions are dominated by early-type stars and considering
their overall size (increasing order) a total of 24 aggregates, 23 complexes, and 3 
super-complexes were found. 
We present 
their coordinates, dimensions, and IR fluxes. Moreover, we correlate their 
positions with known associations, SNRs, and \hii regions and discuss their activity. 
\end{abstract}




\section{INTRODUCTION}

The Magellanic Clouds, being so close to our galaxy, are the most appropriate
example of galaxies in order to study star formation in large scale, and due to 
their proximity, there are data available in various wavelength regimes. 
Moreover, the Magellanic Clouds and the Milky Way form an interacting 
system of galaxies, with tidal forces having significant effects on 
triggering star formation events. 

In a previous paper \citep{livanou}, we studied the connection between 
various tracers of star formation in the LMC, using different wavelength 
regimes. In the present 
paper we conduct an investigation of the SMC, based upon similar data. A
distance of 63~kpc is used throughout; however, owing to the large depth 
of the SMC along the line of sight (\citealt{mathewson}; 
\citealt{hatzidimitriou}), we shall account and correct for this whenever 
necessary. 

Star complexes are the largest and oldest objects among the various size 
stellar groupings, which start from the multiple stars. The younger and 
smaller clusters are always within the older and the larger ones. This 
hierarchy is similar to one observed in the interstellar gas distribution 
and, in fact, is the result of the latter. Sometimes a galaxy hosts some very 
bright complexes, which consist of OB-associations and HII regions. 
They were called long ago the superassociations and represent the local 
starbursts \citep{efremov}. 

Recently, N-body models for tidal interactions showed that the star formation 
rate in the SMC has been strongly enhanced by the tidal encounters with the 
LMC that occurred 1.5~Gyr (14~kpc separation) and 0.2~Gyr (7~kpc separation) 
ago, by a factor of about 3-4. Interestingly, in these simulations the 
peak of the star formation rate occurs 200-300~Myr after the first 
encounter event, so that the star formation rate may still be rising at the 
present epoch, 200~Myr after the second interaction (\citealt{yoshizawa2}). 
It is then reasonable  to attempt tracing the effect of this enhancement 
on the current appearance of the SMC.

\cite{maragoudaki2} have studied the spatial distribution of the SMC stellar 
population according to their age, based on optical data. They concluded that 
the old stellar population shows a rather regular and smooth distribution, 
which is typical for a spheroidal body, while the younger stellar component 
is highly asymmetric and irregular, probably due to the severe impact of its 
close encounter with the LMC some 0.2-0.4~Gyr ago. In particular, the 
wing and the tail structures are already evident at $(3.4-4.0)~\times~
$10$^{8}$~yr, while the new generations of stars appear mainly along the 
northeast-southwest direction forming the bar. The bar becomes prominent at 
an age of about ($1.2-3.0)~\times~$10$^{7}$~yr and hosts all stars with 
age less than 8~$\times$~10$^{6}$~yr, showing how important the recent star 
formation is for the morphology of the SMC and giving us an indication of 
were star-forming, even starburst, regions are expected to be detected.

Star forming regions have been found in both Clouds (\citeauthor{maragoudaki1}
\citeyear{maragoudaki1}, \citeyear{maragoudaki2}) to include smaller regions 
with enhanced star formation activity, supporting the hierarchical scenario 
of large scale star formation process \citep{elmegreen}.
Ranking them by increasing size, we use the terms: aggregates, complexes 
and supercomplexes. Considering the intensity of their star formation activity 
we use the terms: starburst regions and active regions, based on the criteria 
set by \cite{livanou} for the LMC. 
Moreover a comparison with LMC reveals that the relevant frequence of the 
same size structures in the MCs is similar, independently of LMC´s larger 
size. However, star formation is considerably enhanced in the LMC, probably
due to higher total mass, H$_{2}$ abundance and metalicity.
Throughout this work we study the large scale star formation in the SMC.
We investigate the criteria of detection of star forming 
regions and evaluate the intensity of star formation activity in each one and 
compare with LMC.

In Sect.~2 the observational material is presented.
In Sect.~3 we describe how the starforming regions  
are selected in images at 60 and 100~$\mu$m.
We estimate their dimensions, rank them by size 
(aggregates, complexes and super-complexes) and compare with those in
optical and radio data.
In Sect.~4 the intensity of activity of the selected starforming regions
is discussed and their characteristics are given.

\section{OBSERVATIONAL MATERIAL}

\begin{figure}
\includegraphics[angle=0,scale=.75]{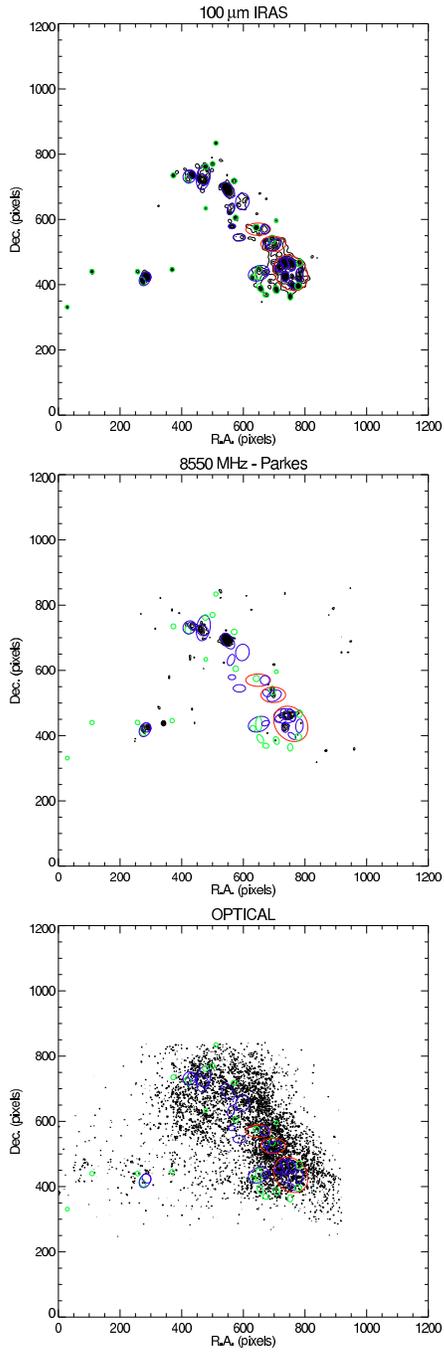}
\caption{Identified star forming regions and their correlation with emission in three
different regions of the spectrum (IRAS 100~$\mu$m, Parkes 8550~MHz and optical 
star counts). Red ellipses indicate the super-complexes,
blue is for complexes, and green for aggregates.}
\label{f1}
\end{figure}

In the optical domain,
we used the direct photographic plates in U, V and I of the SMC, taken with the 1.2-m 
UK Schmidt Telescope. The plates that cover an area of $6\fdg5 \times 6\fdg5$, 
centred at $\alpha = 01^{\rm h}07^{\rm m}$22\fs8, $\delta$ = 
-74\degr44\arcmin00\farcs1\ (J2000), were digitized by the fast measuring 
machine SuperCosmos, providing us with catalogues of detected stars
and their corresponding CMDs \citep{maragoudaki2}. 
Based on these data, we created star count images of main-sequence stars for 
different magnitude slices and pixel sizes, plotting iso-pleth contour maps in 
order to trace stellar concentrations. The optimum result was achieved for a 
750~$\times$~750 pixel image with pixel sizes corresponding to 9.53~pc. 
The mean value $m$ and standard deviation $\sigma$ of the background star 
density were calculated as the average of four relatively empty regions 
in the star count image. For the image of the brightest main sequence stars 
(13\fmag5 $<$ m $<$ 13\fmag83), the minimum 
contour level was set at $m+3\sigma$ above the mean background density and 
the step between different contours was set at 1~$\sigma$.

We also used the 12, 25, 60, and 100~$\mu$m IRAS images by \cite{bontekoe1} 
to produse the infrared isodensity contours of the SMC. They were reconstructed by the 
HIRAS program using the Pyramid Maximum Entropy Method \citep{bontekoe2}. 
The 1024~$\times$~1024 images are centred at $\alpha = 
01^{\rm h}$00$^{\rm m}$54.98$^{\rm s}$ and $\delta$ = 
-72\degr54\arcmin47\arcsec\ (J2000), covering a 4\fdg3~$\times$~4\fdg3\ 
field of view, corresponding to a pixel size of 4.48~pc. For each frame, 
the mean background value $m$ and standard deviation $\sigma$ were 
calculated by the four rather clean corner areas. The plotted contours 
have 1$\sigma$ steps over the $m+3\sigma$ value.  

Ionized gas is prominent in star forming regions and its emission stands out
clearly in the radio domain. Therefore, we used the 8550~MHz radio continuum 
map published by \citep{haynes}. The observation was carried out in 1988 
November using the Parkes 64-m radio telescope over a 400~MHz bandwidth. Sources 
with known flux density and position were also observed and used as 
calibrators. The resulting 201$\times$151 pixel map, centred at $\alpha = 
00^{\rm h}$56$^{\rm m}$ 41\fs4 and $\delta$ = -72\degr43\arcmin47\farcs6\ 
(J2000), covers a 3\fdg5~$\times$~2\fdg5\ field of view, corresponding to a 
pixel size of 28.27~pc. 

\section{DETECTION OF STARFORMING REGIONS}

\subsection{Selection}

The selection of the starforming regions in the SMC was not as straightforward
as that in the LMC \citep{livanou}. The optical image is not the most 
appropriate to identify star-forming regions since the main body of 
the SMC, appearing almost edge-on, absorbs a significant amount of the 
optical light.
The radiation from hot young stars is absorbed by dust grains, which
heat up and re-emit the energy in the infrared.
Especially FIR luminosity is found to be a good indicator of star formation
activity.
It is known that the flux in 60, and 100 $\mu$m characterizes the FIR 
luminosity of a region. The total far-infrared luminosity is:
\begin{equation}\label{eq1_01}
F_{FIR}=C(60/100)\times1.26\times10^{-14}\times(2.58F_{60}+F_{100}) Wm^{-2}
\end{equation} 
where C(60/100) is the color correction dependent on the fitted temperature 
(Cataloged Galaxies and Quasars Observed in the IRAS Survey, 1985)
(\cite{lonsdale}; \cite{helou_88}; \cite{lehnert1}).
Thus the intensity of star formation is well represented from the
images at 60 and 100 $\mu$m.
Almost all of the regions of enhanced infrared luminosity, 
can be identified in all four IRAS images,
but most of them have been found systematically smaller at 12 and 25 $\mu$m, 
so we did not use them for our final results.
We therefore compared the images at 60 and 100 $\mu$m to decide for the final 
selection and size of the adopted regions.
All selected regions are found in both maps with similar sizes (less than 10\%
difference), thus we could choose either to reveal the borders of the starforming regions.
We selected areas with increased 60 and 100-$\mu$m infrared flux, 
which show emission in all IRAS 
images and resulted in 50 star forming regions (Fig.\ref{f1} top).
The calculated emission of each area in the four IRAS wavebands can be seen in
Table~\ref{t2}.

\subsection{Sizes}

For the size classification of the starforming regions we adopted the values given by 
\cite{maragoudaki1}. {\it Aggregates} are groupings with sizes between 150 
and 300~pc, {\it complexes} range between 300 and 1000~pc and 
{\it super-complexes} exceed 1000~pc in diameter, which in our case is set as
the major axes of the fitted ellipses. It has to be emphasized that all large
structures are found to include smaller. In an attempt to estimate 
the sizes of these areas, ellipses were fitted to the structures in the 
100-$\mu$m IRAS image, such as to resemble the structure of each area in 
the best possible way. However, the edge-on orientation of the SMC required 
projection corrections to be taken into account. After applying these 
corrections, 24 aggregates, 23 complexes, and 3 super-complexes were revealed. 
Fig.~\ref{f1} shows these regions overplotted on the maps at 100~$\mu$m, 
8550~MHz and the optical, with green ellipses indicating aggregates, 
blue corresponding to complexes, and red depicting super-complexes.   
From the correlation with the optical image of the brightest main sequence
stars, we infer the age of the stellar content. The age of this stellar 
population is younger than 8$\times$10$^{6}$~yr (\citealt{maragoudaki2}),
giving evidence of being a result of the most recent encounter of SMC and LMC
about $(0.9-2)~\times~10^{8}$ years ago (\citealt{gardiner}).

\section{ACTIVITY OF STARFORMING REGIONS}

\begin{figure}
\includegraphics[angle=0,scale=.40]{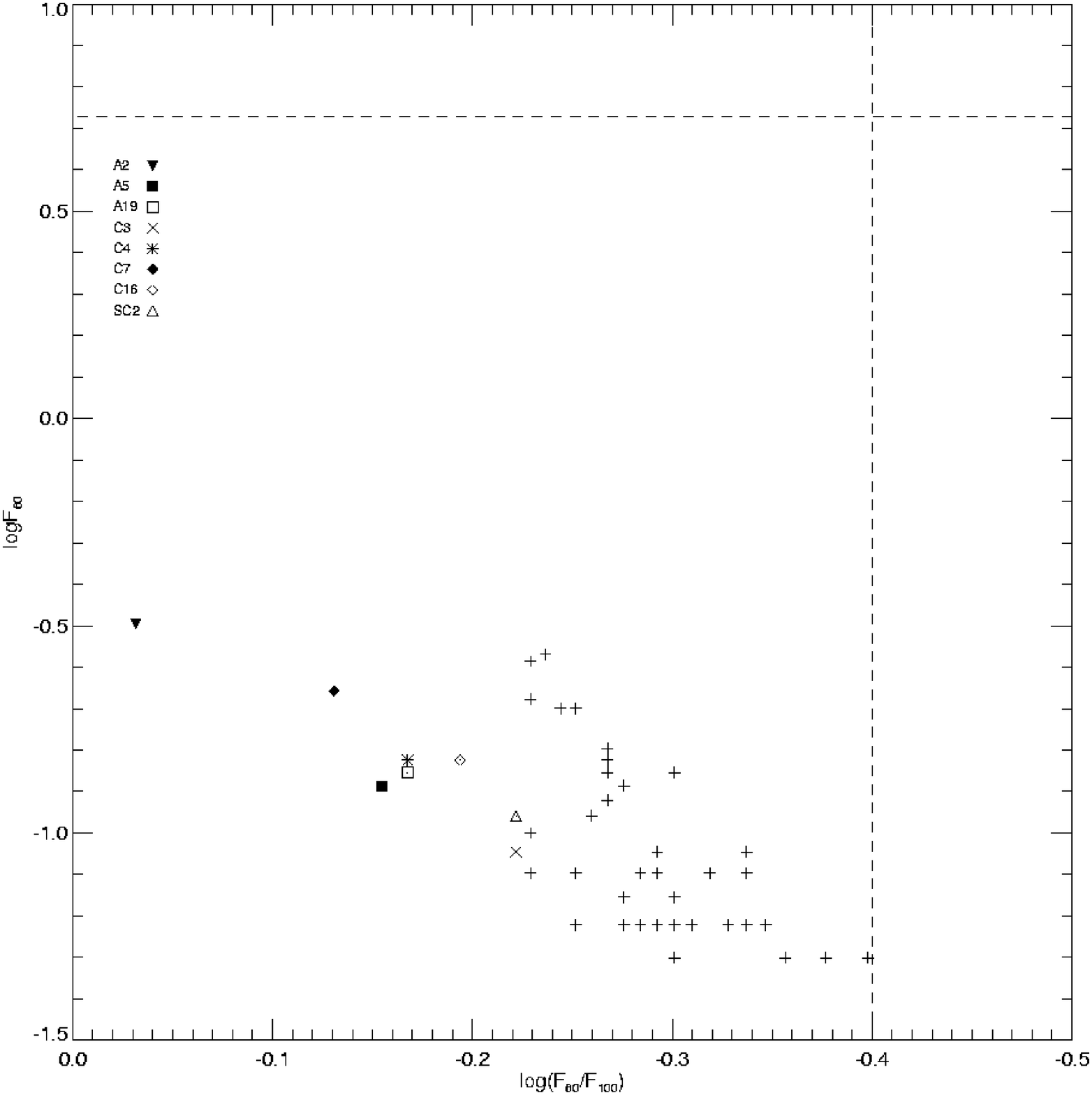}
\caption{Flux at 60~$\mu$m versus the colour log(F$_{60}$/F$_{100}$)
for the SMC detected regions. The eight regions indicated by different symbols,
are more active in star formation than the rest.}
\label{f2}
\end{figure}

\cite{lehnert1} found that IR-``warm'' (F$_{60}$/F$_{100} \ge$ 0.4) and 
IR-``bright'' (F$_{60} \ge$ 5.4~Jy) galaxies show strong evidence of forming 
massive stars at unusually high rates. Thus if any of the detected star 
forming regions was found with F$_{60}$/F$_{100} \ge$ 0.4 and 
F$_{60} \ge$ 5.4~Jy, it could be considered as a region of enhanced star formation activity.
It could even be characterized as starburst if it also had excess of 8550~MHz
emission \citep{livanou}.
The correlation of the groupings with the \cite{lehnert1} criterion can be 
seen in Fig.~\ref{f2}. It is apparent that 8 regions are clearly 
separated relatively to the rest, occupying the left part of the diagram. 
However, none of the groupings have F$_{60}$ above the 5.4~Jy threshold. 

The stage of evolution of the regions can be estimated by the 8550~MHz map, 
where enhanced emission reveals areas of on-going star formation, since it reflects 
mainly thermal emission from ionized gas. Col.~10 of Table~\ref{t1} indicates the 
existence or not of 8550~MHz emission, after correlation of the detected 
star forming regions with the 8550~MHz map. The initial radio map had different pixel 
size and field of view from the IRAS images. 
A scaling was performed in all images in order 
to have the same characteristics, therefore the final images are identical 
in terms of their sizes and angular resolution. However, there are areas 
that are not commonly contained in all the images used here, since the 
observations in the four different wavebands (radio continuum, FIR, 
optical) have different coverages; such areas are indicated with `n/a' in 
Table~\ref{t1} and \ref{t3}. Apart from the stellar grouping \emph{A2}, 
which is outside the radio continuum map, the other 7 regions are strongly 
correlated with the 8550-MHz~emission. 

Although all eight groupings are correlated with areas where very bright stars
are located (see Fig.~\ref{f1} - optical), none of these eight regions 
exhibit any increased formation of high-mass stars according to the 
\cite{lehnert1} criterion. Contrary to the LMC, where 13 such regions were 
found, ``starburst regions'' in the SMC were not identified. The 50 
stellar groupings listed here are ``active complexes'' with clear 
indications for eight of them showing more intense activity of star formation.

\begin{figure}
\includegraphics[angle=0,scale=.75]{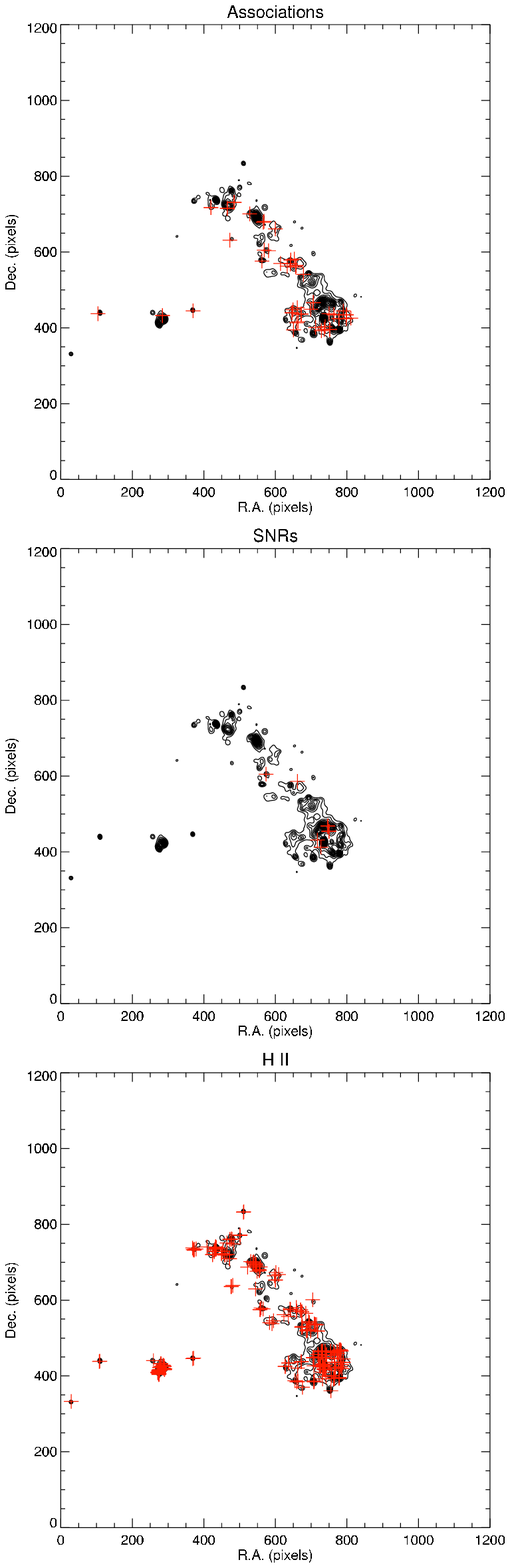}
\caption{Correlated associations, SNR, and \hii regions plotted over the 
100~$\mu$m IRAS image.}
\label{f3}
\end{figure}

Finally, starbursts are expected to 
have a larger amount of SNRs, \hii regions, and stellar associations than normal 
galaxies. Fig.~\ref{f3} shows the distribution of stellar associations, 
SNRs and \hii regions, plotted over the IRAS 100-$\mu$m image.    
Results can be seen in Cols.~7, 8 and 9 of Table~\ref{t1}.
The catalogue compiled by 
\cite{bica} is used to cross match the areas of interest with these objects. 
The detected starforming regions obviously contain a larger number of SNRs, 
\hii regions, and stellar associations than their environment. This fact 
strengthens our claim that star formation is enhanced in these areas.

\section{COMPARISON OF THE LMC WITH THE SMC}

\begin{figure}
\includegraphics[angle=0,scale=.50]{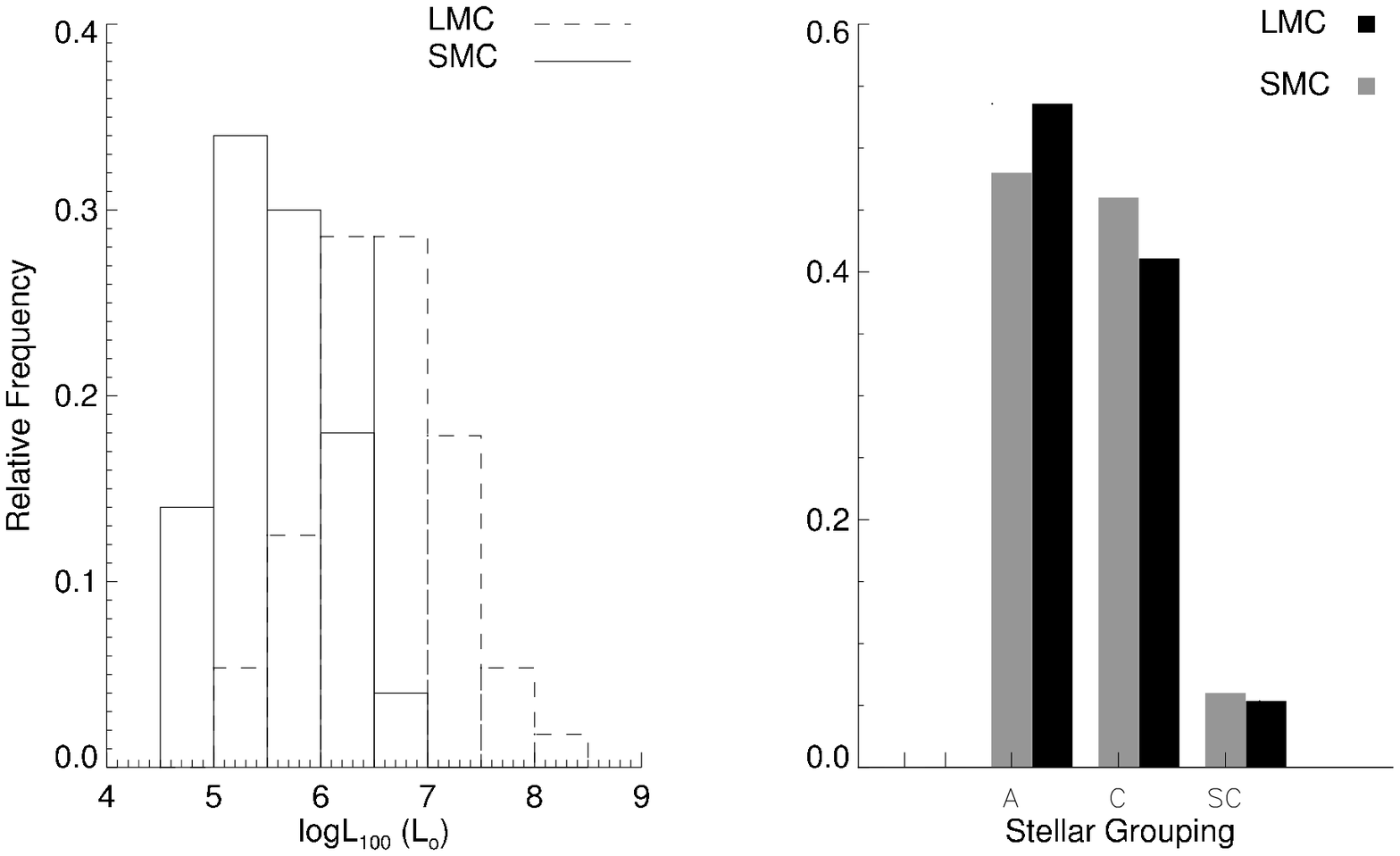}
\caption{Relative frequency 
distribution of star forming regions with respect to their 100~$\mu$m-luminosity, 
in units of \Lsun (left) and relative frequency distribution of the three 
kinds of star forming regions (right) for LMC and SMC.}
\label{f4}
\end{figure}

Fig.~\ref{f4} shows a comparison between the two Magellanic Clouds. The left panel 
shows the frequency distribution of the IR luminosities at 100~$\mu$m, with
solid lines for the SMC and dashed for the LMC. The right panel shows the
frequency distribution for each kind of stellar grouping identified in each
galaxy, where black bars represent the LMC and grey the SMC. 
LMC is almost face-on so there are no intrinsic differences in the luminosity 
due to distance. For the SMC corrections due to depth have to be considered.

\cite{mathewson} studied the distances of 61 Cepheids along the SMC bar,
revealing a three dimensional prespective for the SMC. The distribution of
these Cepheids showed that the northeastern section of the bar is 10-15~kpc
closer than the southern section.
Thus, luminosities in the northeast, 
the central and the southwest part
of the bar may vary up to 12.3, 6.25, and 9.75 \% respectively. However, in
some extreme cases this factor may extend up to 50 \%.  
  
It appears 
that the relative frequencies of aggregates, complexes, and super-complexes in 
both galaxies are almost the same, while the star-forming activity differs
significantly. The histograms of their 100-$\mu$m luminosities (adopted at 63
kpc) were fitted with Gaussians such as to reveal similarities in these 
distributions. Apart from a shift of their centres, the Gaussians reflect 
similar properties. The best fit for the SMC produces a Gaussian centred 
at logL$_{100}$=5.51 (in units of \Lsun), with an amplitude of 0.349$\pm$0.015 
and a standard deviation of 0.60$\pm$0.03, while for the LMC the center is 
located at logL$_{100}$=6.56, with an amplitude of 0.307$\pm$0.025 and a 
standard deviation of 0.648$\pm$0.024. Within the range defined by the errors, 
shapes of these Gaussians can be considered as identical. The shift is not 
significantly affected by the depth, as explained above and can be a result
of the known difference in the colour-corrected 
flux F$_{100}$ for the SMC is 15.51$\times$10$^{3}$~Jy and 
190.6$\times$10$^{3}$~Jy for the LMC \citep{rice}.

\section{CONCLUSIONS}

We have identified the large scale star-forming regions in the SMC in order to investigate 
its activity. This study is associated to the past interaction history 
of the LMC and SMC, which had a close encounter about $(0.9-2)~\times~10^{8}$ 
years ago. For the identification of the star-forming regions a selection was 
based upon an increased 60 and 100-$\mu$m infrared flux 
and then compared to the optical and radio image. Ranking them by increasing order 
of size, we detected 24 aggregates, 23 complexes, and 3 
super-complexes. It has to be emphasized that all large structures
include smaller ones, supporting the hierarchical scenario of star formation in large 
scales.

All of the identified regions are dominated by early-type stars. 
These stars are found to be younger than $\sim$8$\times$10$^{8}$ years old and 
thus are believed to be formed as a consequence of the LMC-SMC close encounter.

In order to classify these regions in terms of their activity we compare 
their IR fluxes with those of other galaxies, in particular starburst galaxies.
We find none of the star forming regions detected here to fulfill the criteria 
of starburst regions contrary to what is observed in the LMC. However 8 of the 
star forming regions are more active than the rest.

By inspecting optical and radio images of the SMC, we have assessed any 
correlation with known associations, SNRs and \hii regions. The southern-most 
part of the bar of the SMC appears to be the most diverse in terms of star
formation activity, which is probably connected with its large depth along 
the line of sight. For the northern part of the bar we conjecture that star 
formation has started in the recent past, mainly inferred from the lack of SNRs 
and the low number of stellar associations.

If this activity in both galaxies is considered to be triggered, during their 
recent close encounter, the intensity of this activity is evidently much higher
in the LMC than in the SMC. Could this reflect the difference in the total mass,
in gas or metallicity between these galaxies? Was the gas removed from the less 
massive galaxy (SMC) more easily preventing high star formation activity?

\begin{table*}
\tiny{
\begin{center}
\caption{Star forming regions of the SMC and their properties.\label{t1}}
\begin{tabular}{lccccccccc}
\tableline\tableline
Star forming  & RA (J2000) & DEC (J2000)  & RA       & Dec     & Dimension  & No. of        & No. of   & No. of HII    & 8550        \\
region        & h   m   s  & deg  m  s    & Pixels   & Pixels  & (pc)       & Associations  & SNRs     & regions        & MHz         \\
\tableline
 A1 &    1:29:27 &   -73:33:50 &  29 & 331 &  160.52 &  0 &  0 &  1  &   n/a    \\
 A2 &    1:24:09 &   -73:08:54 & 109 & 440 &  187.27 &  1 &  0 &  2  &   n/a    \\
 A3 &    1:15:39 &   -73:11:48 & 257 & 440 &  191.90 &  0 &  0 &  1  &   -      \\
 A4 &    1:14:50 &   -73:19:17 & 273 & 411 &  266.62 &  0 &  0 & 12  &  $\surd$ \\
 C1 &    1:14:21 &   -73:17:39 & 281 & 418 &  505.38 &  1 &  0 & 23  &  $\surd$ \\
 C2 &    1:14:02 &   -73:15:58 & 286 & 425 &  377.06 &  1 &  0 & 12  &  $\surd$ \\
 A5 &    1:09:11 &   -73:11:39 & 369 & 447 &  187.27 &  1 &  0 &  2  &  $\surd$ \\
 A6 &    1:08:26 &   -71:59:28 & 373 & 735 &  214.02 &  0 &  0 &  5  &   -      \\
 A7 &    1:05:55 &   -72:02:46 & 420 & 723 &  257.37 &  1 &  0 &  2  &  $\surd$ \\
 C3 &    1:05:38 &   -72:00:33 & 425 & 732 &  551.47 &  1 &  0 &  9  &  $\surd$ \\
 C4 &    1:05:12 &   -71:59:20 & 433 & 737 &  338.64 &  0 &  0 &  7  &  $\surd$ \\
 C5 &    1:03:23 &   -72:02:57 & 467 & 723 &  508.85 &  2 &  0 &  5  &  $\surd$ \\
 C6 &    1:03:13 &   -72:01:12 & 470 & 730 &  619.09 &  2 &  0 & 14  &  $\surd$ \\
 A8 &    1:02:49 &   -71:53:13 & 477 & 762 &  240.78 &  0 &  0 &  4  &  $\surd$ \\
 A9 &    1:02:49 &   -72:25:13 & 478 & 634 &  160.52 &  1 &  0 &  3  &   -      \\
A10 &    1:01:35 &   -71:51:15 & 500 & 770 &  214.02 &  0 &  0 &  2  &   -      \\
A11 &    1:00:60 &   -71:35:15 & 511 & 834 &  187.27 &  0 &  0 &  2  &   -      \\
 C7 &    0:58:53 &   -72:11:12 & 550 & 690 &  588.40 &  3 &  0 & 13  &  $\surd$ \\
 C8 &    0:58:18 &   -72:25:41 & 560 & 632 &  331.69 &  0 &  0 &  1  &   -      \\
 C9 &    0:58:06 &   -72:38:55 & 563 & 579 &  321.04 &  1 &  0 &  4  &   -      \\
A12 &    0:57:48 &   -72:04:09 & 570 & 718 &  241.47 &  0 &  0 &  0  &   -      \\
A13 &    0:57:27 &   -72:32:23 & 575 & 605 &  240.78 &  2 &  1 &  0  &   -      \\
C10 &    0:56:43 &   -72:47:20 & 587 & 545 &  535.06 &  0 &  0 &  3  &   -      \\
C11 &    0:56:17 &   -72:19:47 & 597 & 655 &  591.74 &  1 &  0 &  5  &   -      \\
A14 &    0:54:02 &   -73:17:49 & 631 & 422 &  242.92 &  0 &  0 &  1  &   -      \\
A15 &    0:53:40 &   -72:39:30 & 642 & 575 &  267.53 &  2 &  0 &  2  &   -      \\
SC1 &    0:53:20 &   -72:40:42 & 648 & 570 & 1070.12 &  7 &  1 &  9  &  $\surd$ \\
C12 &    0:52:58 &   -73:14:25 & 650 & 435 &  902.74 &  6 &  0 &  4  &   -      \\
A16 &    0:52:58 &   -73:14:10 & 650 & 436 &  267.53 &  2 &  0 &  2  &   -      \\
A17 &    0:52:35 &   -73:25:37 & 655 & 390 &  292.28 &  1 &  0 &  3  &   -      \\
C13 &    0:52:06 &   -72:40:31 & 670 & 570 &  401.30 &  1 &  0 &  3  &   -      \\
C14 &    0:51:42 &   -73:13:28 & 672 & 438 &  329.35 &  1 &  0 &  1  &   -      \\
A18 &    0:51:29 &   -73:30:42 & 673 & 369 &  259.47 &  0 &  0 &  1  &   -      \\
C15 &    0:51:44 &   -72:50:27 & 675 & 530 &  333.77 &  0 &  0 &  2  &   -      \\
A19 &    0:50:41 &   -72:48:16 & 694 & 538 &  267.53 &  0 &  0 &  2  &  $\surd$ \\
SC2 &    0:50:29 &   -72:51:29 & 697 & 525 & 1070.12 &  1 &  0 & 12  &  $\surd$ \\
C16 &    0:50:18 &   -72:52:12 & 700 & 522 &  615.87 &  0 &  0 &  7  &  $\surd$ \\
A20 &    0:50:06 &   -72:33:38 & 707 & 596 &  160.52 &  0 &  0 &  1  &   -      \\
A21 &    0:49:32 &   -73:26:20 & 707 & 385 &  253.52 &  0 &  0 &  3  &  $\surd$ \\
C17 &    0:49:12 &   -73:09:14 & 716 & 453 &  354.85 &  0 &  0 &  4  &  $\surd$ \\
C18 &    0:48:13 &   -73:06:01 & 734 & 465 &  347.19 &  0 &  1 &  7  &  $\surd$ \\
C19 &    0:48:08 &   -73:07:15 & 735 & 460 &  963.11 &  1 &  2 & 17  &  $\surd$ \\
C20 &    0:47:58 &   -73:15:43 & 736 & 426 &  321.04 &  0 &  0 & 12  &  $\surd$ \\
A22 &    0:46:50 &   -73:30:58 & 752 & 364 &  214.03 &  0 &  0 &  1  &   -      \\
SC3 &    0:46:58 &   -73:12:58 & 754 & 436 & 1479.86 & 14 &  4 & 56  &  $\surd$ \\
C21 &    0:46:57 &   -73:06:12 & 756 & 463 &  353.92 &  0 &  2 &  8  &  $\surd$ \\
C22 &    0:46:36 &   -73:22:09 & 758 & 399 &  328.43 &  1 &  0 &  3  &   -      \\
A23 &    0:45:26 &   -73:22:35 & 778 & 396 &  294.28 &  0 &  0 &  2  &  $\surd$ \\
A24 &    0:45:28 &   -73:04:48 & 782 & 467 &  263.93 &  0 &  0 &  7  &  $\surd$ \\
C23 &    0:45:20 &   -73:14:02 & 782 & 430 &  319.95 &  4 &  0 &  8  &  $\surd$ \\
\tableline
\end{tabular}
\tablecomments{
SMC star forming regions: Col.~1 contains the adopted 
identification name. \emph{A} denotes aggregate, \emph{C} complex, and 
\emph{SC} super-complex. Cols.~2 and 3 contain the right ascension and 
declination of the ``center'' of each star forming region, that are also 
given in pixels in Cols.~4 and 5. Col.~6 gives the dimension of each 
star forming region, while Cols.~7, 8, and 9 list the number of associations, 
SNRs, and \hii regions found in each region. Col.~10 list the 
8550-MHz flux. Areas that have not been 
identified due to incomplete coverage of the field of view are marked as 
`n/a' in the table.}
\end{center}
}
\end{table*}

\begin{table*}
\tiny{
\begin{center}
\caption{List of the derived IRAS fluxes for the SMC star forming regions.\label{t2}}
\begin{tabular}{ccccccccccc}
\tableline\tableline
Star forming& f12   & L12  & f25  & L25  & f60  & L60 & f100  & L100 & $F_{60}/F_{100}$ &
$F_{60}$\\
region & & & & & & & & & & \\
\tableline
 A1 & 0.11 & 2.01e+03 & 1.02 & 8.97e+03 & 11.45 & 4.21e+04 & 24.26 & 5.35e+04 & 0.47 & 0.06\\
 A2 & 1.89 & 4.73e+04 &13.20 & 1.58e+05 & 59.71 & 2.98e+05 & 63.92 & 1.92e+05 & 0.93 & 0.32\\
 A3 & 0.22 & 6.41e+03 & 0.66 & 9.08e+03 &  8.75 & 5.00e+04 & 21.81 & 7.48e+04 & 0.40 & 0.05\\
 A4 & 0.34 & 2.21e+04 & 2.82 & 8.70e+04 & 26.37 & 3.39e+05 & 48.40 & 3.73e+05 & 0.54 & 0.14\\
 C1 & 0.42 & 8.77e+04 & 2.37 & 2.37e+05 & 24.12 & 1.00e+06 & 45.47 & 1.14e+06 & 0.53 & 0.13\\
 C2 & 0.68 & 7.20e+04 & 3.10 & 1.58e+05 & 30.68 & 6.51e+05 & 57.09 & 7.27e+05 & 0.54 & 0.16\\
 A5 & 0.40 & 9.89e+03 & 4.06 & 4.87e+04 & 24.17 & 1.21e+05 & 34.62 & 1.04e+05 & 0.70 & 0.13\\
 A6 & 0.17 & 5.42e+03 & 0.88 & 1.38e+04 & 13.27 & 8.66e+04 & 24.94 & 9.77e+04 & 0.53 & 0.07\\
 A7 & 0.21 & 1.35e+04 & 0.56 & 1.74e+04 & 11.76 & 1.51e+05 & 23.22 & 1.79e+05 & 0.51 & 0.06\\
 C3 & 0.34 & 7.66e+04 & 1.69 & 1.82e+05 & 16.66 & 7.48e+05 & 27.76 & 7.48e+05 & 0.60 & 0.09\\
 C4 & 0.67 & 5.29e+04 & 4.06 & 1.53e+05 & 29.23 & 4.59e+05 & 43.10 & 4.06e+05 & 0.68 & 0.15\\
 C5 & 0.40 & 6.89e+04 & 1.42 & 1.17e+05 & 21.77 & 7.46e+05 & 40.28 & 8.28e+05 & 0.54 & 0.12\\
 C6 & 0.29 & 1.35e+05 & 0.91 & 2.05e+05 & 15.19 & 1.43e+06 & 29.79 & 1.68e+06 & 0.51 & 0.08\\
 A8 & 0.41 & 1.69e+04 & 1.27 & 2.51e+04 & 17.27 & 1.43e+05 & 37.49 & 1.86e+05 & 0.46 & 0.09\\
 A9 & 0.09 & 1.73e+03 & 0.34 & 2.95e+03 &  9.68 & 3.56e+04 & 19.17 & 4.22e+04 & 0.50 & 0.05\\
A10 & 0.14 & 4.59e+03 & 0.98 & 1.54e+04 & 11.20 & 7.31e+04 & 21.63 & 8.47e+04 & 0.52 & 0.06\\
A11 & 0.27 & 6.85e+03 & 1.39 & 1.66e+04 & 14.48 & 7.24e+04 & 25.92 & 7.78e+04 & 0.56 & 0.08\\
 C7 & 0.60 & 1.50e+05 & 4.30 & 5.19e+05 & 40.93 & 2.06e+06 & 55.01 & 1.66e+06 & 0.74 & 0.22\\
 C8 & 0.22 & 2.34e+04 & 1.14 & 5.86e+04 & 11.91 & 2.54e+05 & 24.17 & 3.09e+05 & 0.49 & 0.06\\
 C9 & 0.18 & 8.95e+03 & 0.82 & 1.92e+04 & 13.24 & 1.30e+05 & 26.57 & 1.56e+05 & 0.50 & 0.07\\
A12 & 0.16 & 6.65e+03 & 0.76 & 1.50e+04 &  9.83 & 8.10e+04 & 22.53 & 1.11e+05 & 0.44 & 0.05\\
A13 & 0.20 & 8.14e+03 & 0.93 & 1.85e+04 & 11.83 & 9.77e+04 & 22.14 & 1.10e+05 & 0.53 & 0.06\\
C10 & 0.18 & 2.14e+04 & 0.54 & 3.20e+04 & 11.02 & 2.70e+05 & 22.24 & 3.27e+05 & 0.50 & 0.06\\
C11 & 0.17 & 5.23e+04 & 0.41 & 5.90e+04 & 10.09 & 6.11e+05 & 20.36 & 7.40e+05 & 0.50 & 0.05\\
A14 & 0.24 & 1.12e+04 & 0.99 & 2.19e+04 & 11.50 & 1.06e+05 & 24.49 & 1.35e+05 & 0.47 & 0.06\\
A15 & 0.21 & 1.08e+04 & 0.68 & 1.66e+04 & 14.94 & 1.52e+05 & 28.87 & 1.77e+05 & 0.52 & 0.08\\
SC1 & 0.17 & 7.12e+04 & 0.52 & 1.01e+05 & 11.80 & 9.63e+05 & 21.09 & 1.03e+06 & 0.56 & 0.06\\
C12 & 0.22 & 9.49e+04 & 0.51 & 1.05e+05 & 11.44 & 9.81e+05 & 24.89 & 1.28e+06 & 0.46 & 0.06\\
A16 & 0.27 & 3.22e+04 & 0.58 & 3.28e+04 & 14.62 & 3.43e+05 & 31.95 & 4.50e+05 & 0.46 & 0.08\\
A17 & 0.16 & 1.12e+04 & 0.64 & 2.12e+04 & 11.49 & 1.58e+05 & 25.41 & 2.10e+05 & 0.45 & 0.06\\
C13 & 0.25 & 2.81e+04 & 0.61 & 3.36e+04 & 14.58 & 3.35e+05 & 24.74 & 3.41e+05 & 0.59 & 0.08\\
C14 & 0.26 & 1.37e+04 & 0.76 & 1.93e+04 & 16.25 & 1.72e+05 & 32.13 & 2.05e+05 & 0.51 & 0.09\\
A18 & 0.05 & 2.22e+03 & 0.28 & 5.56e+03 &  9.14 & 7.46e+04 & 21.86 & 1.07e+05 & 0.42 & 0.05\\
C15 & 0.20 & 1.43e+04 & 0.81 & 2.79e+04 & 19.34 & 2.76e+05 & 32.89 & 2.82e+05 & 0.59 & 0.10\\
A19 & 0.28 & 2.02e+04 & 1.61 & 5.50e+04 & 26.82 & 3.83e+05 & 39.19 & 3.36e+05 & 0.68 & 0.14\\
SC2 & 0.19 & 9.75e+04 & 0.91 & 2.22e+05 & 20.38 & 2.08e+06 & 33.72 & 2.06e+06 & 0.60 & 0.11\\
C16 & 0.24 & 4.81e+04 & 1.27 & 1.25e+05 & 27.76 & 1.13e+06 & 43.07 & 1.05e+06 & 0.64 & 0.15\\
A20 & 0.09 & 1.59e+03 & 0.41 & 3.63e+03 & 11.96 & 4.39e+04 & 21.41 & 4.72e+04 & 0.56 & 0.06\\
A21 & 0.41 & 2.45e+04 & 1.04 & 2.97e+04 & 16.35 & 1.95e+05 & 31.89 & 2.28e+05 & 0.51 & 0.09\\
C17 & 0.63 & 5.29e+04 & 2.25 & 9.09e+04 & 37.99 & 6.39e+05 & 67.47 & 6.82e+05 & 0.56 & 0.20\\
C18 & 0.80 & 9.25e+04 & 2.68 & 1.49e+05 & 50.88 & 1.18e+06 & 88.26 & 1.23e+06 & 0.58 & 0.27\\
C19 & 0.60 & 2.66e+05 & 1.91 & 4.04e+05 & 37.30 & 3.29e+06 & 65.70 & 3.47e+06 & 0.57 & 0.20\\
C20 & 1.05 & 1.09e+05 & 4.36 & 2.18e+05 & 49.94 & 1.04e+06 & 83.98 & 1.05e+06 & 0.59 & 0.26\\
A22 & 0.34 & 1.68e+04 & 0.86 & 2.01e+04 & 14.38 & 1.41e+05 & 30.15 & 1.77e+05 & 0.48 & 0.08\\
SC3 & 0.46 & 7.29e+05 & 1.46 & 1.11e+06 & 25.93 & 8.20e+06 & 47.70 & 9.05e+06 & 0.54 & 0.14\\
C21 & 0.65 & 7.97e+04 & 2.00 & 1.18e+05 & 39.43 & 9.74e+05 & 67.38 & 9.98e+05 & 0.59 & 0.21\\
C22 & 0.68 & 4.15e+04 & 1.59 & 4.68e+04 & 29.26 & 3.58e+05 & 54.04 & 3.97e+05 & 0.54 & 0.15\\
A23 & 0.61 & 3.77e+04 & 1.88 & 5.57e+04 & 26.61 & 3.28e+05 & 53.42 & 3.96e+05 & 0.50 & 0.14\\
A24 & 0.35 & 2.13e+04 & 1.07 & 3.13e+04 & 20.69 & 2.53e+05 & 37.33 & 2.74e+05 & 0.55 & 0.11\\
C23 & 0.55 & 7.36e+04 & 1.76 & 1.14e+05 & 26.83 & 7.23e+05 & 49.98 & 8.08e+05 & 0.54 & 0.14\\
\tableline
\end{tabular}
\tablecomments{Derived IRAS fluxes (MJy~sr$^{-1}$), luminosities (in 
units of \Lsun) of the identified starforming region at 12, 25, 60, and 
100~$\mu m$, the flux ratio $F_{60}/F_{100}$ and the flux at 60~$\mu m$ 
in Jy per pixel.}
\end{center}
}
\end{table*}

\begin{table*}
\tiny{
\begin{center}
\caption{Estimations of uncertenties for the IRAS fluxes involved in diagrams. Indicative values for 8550~MHz emission are also given.\label{t3}}
\begin{tabular}{lccr}
\tableline\tableline
Star forming       & error         & error                 & 8550   \\
region       & Log(F$_{60}$)  & Log($F_{60}$/F$_{100}$)    & MHz    \\
\tableline
 A1 &    0.05 &    0.07 &      n/a   \\
 A2 &    0.01 &    0.02 &      n/a   \\
 A3 &    0.07 &    0.08 &      0.00  \\
 A4 &    0.02 &    0.03 &     30.85  \\
 C1 &    0.03 &    0.03 &     42.60  \\
 C2 &    0.02 &    0.03 &     66.63  \\
 A5 &    0.03 &    0.04 &      0.00  \\
 A6 &    0.05 &    0.06 &      0.00  \\
 A7 &    0.05 &    0.07 &      0.52  \\
 C3 &    0.04 &    0.05 &     27.70  \\
 C4 &    0.02 &    0.03 &     66.56  \\
 C5 &    0.03 &    0.04 &     68.71  \\
 C6 &    0.04 &    0.05 &     28.67  \\
 A8 &    0.04 &    0.04 &      0.23  \\
 A9 &    0.06 &    0.08 &      0.00  \\
A10 &    0.06 &    0.07 &      0.00  \\
A11 &    0.04 &    0.06 &      0.00  \\
 C7 &    0.02 &    0.02 &    148.33  \\
 C8 &    0.05 &    0.07 &      0.00  \\
 C9 &    0.05 &    0.06 &      0.00  \\
A12 &    0.06 &    0.08 &      0.00  \\
A13 &    0.05 &    0.07 &      0.00  \\
C10 &    0.06 &    0.07 &      0.00  \\
C11 &    0.06 &    0.08 &      0.00  \\
A14 &    0.05 &    0.07 &      0.00  \\
A15 &    0.04 &    0.05 &      0.00  \\
SC1 &    0.05 &    0.07 &      0.05  \\
C12 &    0.05 &    0.07 &      0.00  \\
A16 &    0.04 &    0.05 &      0.00  \\
A17 &    0.05 &    0.07 &      0.00  \\
C13 &    0.04 &    0.06 &      0.00  \\
C14 &    0.04 &    0.05 &      0.00  \\
A18 &    0.07 &    0.08 &      0.00  \\
C15 &    0.03 &    0.04 &      0.00  \\
A19 &    0.02 &    0.04 &     48.39  \\
SC2 &    0.03 &    0.04 &      8.72  \\
C16 &    0.02 &    0.03 &     12.95  \\
A20 &    0.05 &    0.07 &      0.00  \\
A21 &    0.04 &    0.05 &      1.31  \\
C17 &    0.02 &    0.02 &     15.78  \\
C18 &    0.01 &    0.02 &     73.83  \\
C19 &    0.02 &    0.02 &     43.19  \\
C20 &    0.01 &    0.02 &     56.32  \\
A22 &    0.04 &    0.05 &      0.00  \\
SC3 &    0.02 &    0.03 &     16.29  \\
C21 &    0.02 &    0.02 &     72.48  \\
C22 &    0.02 &    0.03 &      0.00  \\
A23 &    0.02 &    0.03 &      0.46  \\
A24 &    0.03 &    0.04 &     12.14  \\
C23 &    0.02 &    0.03 &      0.52  \\
\tableline
\end{tabular}
\tablecomments{
SMC star forming regions: Col.~1 contains the adopted 
identification name. 
Cols.~2 and 3 contain errors for Log(F60) and
Log(60/100), while Col.~4 gives the 8550~MHz emission.}
\end{center}
}
\end{table*}

\section{ACKNOWLEDGMENTS}

The authors wish to thank the General Secretariat of Research and Technology 
for financial support. The project is co-funded by the European Social Fund
and National Resources-(EPEAEK II) PYTHAGORAS II. U. Klein is very grateful 
to the Department of Astrophysics, Astronomy \& Mechanics at the University 
of Athens for the kind hospitality.

\end{document}